# Multiple Plasma Stream Electromechanical Model for Pulsed Plasma Thrusters

Kaartikey Misra[1]

*Manipal Institute of Technology, Manipal, India*

**Abstract**

Conventional snowplow/1-D models assume the existence of single completely ionized plasma sheet in the discharge chamber of pulsed plasma thrusters (PPTs). However, extensive experimental analysis and observations suggest the presence of multiple plasma streams and partial ionization of the propellant gas. A multi-layered plasma-neutral mass sheet one-dimensional numerical model is developed for PPTs in the present paper. The model accounts for the secondary plasma sheets generated due to the crowbar breakdowns as well as the neutral gas generated after the main discharge. The model accurately calculates the position of the plasma and neutral mass sheets, mean exit velocity, thrust delivered and the mass of individual plasma sheet developed in PPTs. The model is validated and discussed for two different ablative pulsed plasma thrusters. In the last section, the plasma sheet model and electrical circuit model are also validated for a liquid propellant PPT developed at the University of Tokyo. Results show that the multi-sheet assumption provides an accurate prediction for the discharge current and performance parameters. The multiple sheet assumption model accounts for most of the experimentally observed phenomena's which the conventional slug models tend to ignore.

## 1. Introduction

Pulsed plasma thrusters (PPTs[2]) are one of the simplest and cheapest form of electric propulsion systems for small spacecrafts. The low power requirements ($< 20W$), simple design and robustness are some of the key factors making PPTs attractive and viable option for space application [1][2]. Due to the relatively simple and cheap design, PPTs have been investigated both experimentally and numerically across various research organizations. Figure 1 depicts the schematic for the operation of a solid propellant PPT.

One key drawback faced by PPTs till date is their low efficiency (typically $< 10\%$) [1][3]. Poor propellant utilization due to the generation of heavy neutral mass is the major reason identified for this low efficiency [3][4].

Experimental results have concluded that only a small fraction of propellant mass shot per discharge is completely ionized and accelerated electromagnetically, whereas, a great fraction of mass shot is composed of macro neutral gas particles and accelerated to a relatively low thermal exit velocity [4][5][6]. As an example, for the Lincoln Experimental Satellite thruster (LES-6), the charge accumulation on the measuring Faraday cup showed that only approximately $1\mu g$ of the total $10\mu g$ mass shot was ionized [7][9].

Another observation made in the past is the existence of multiple plasma sheet layers in the discharge chamber [10]. For an under-damped discharge current waveform, a secondary "crowbar" breakdown is initiated just before the current reversal at the position of least inductance in the discharge channel. This effectively causes the short circuiting of the initial accelerating plasma stream from the external circuitry [1][2]. Crowbar discharges are self triggering and once the discharge voltage reaches a maximum/minimum value a secondary plasma sheet is formed near the propellant face, these processes were investigated extensively in the past at the NASA Lewis Research center [11][12]. Recent studies using emission spectrograms of the PPT discharge also suggests the existence of two plasmoids and also a slow moving neutral mass layer [13]. Numerical studies are an accurate way of understanding and explaining various physical processes observed through experimental analysis.

[1]Undergraduate student, Manipal Institute of Technology, Manipal-576104, India

[2]Present paper uses the term ablative PPT and PPT interchangeably.

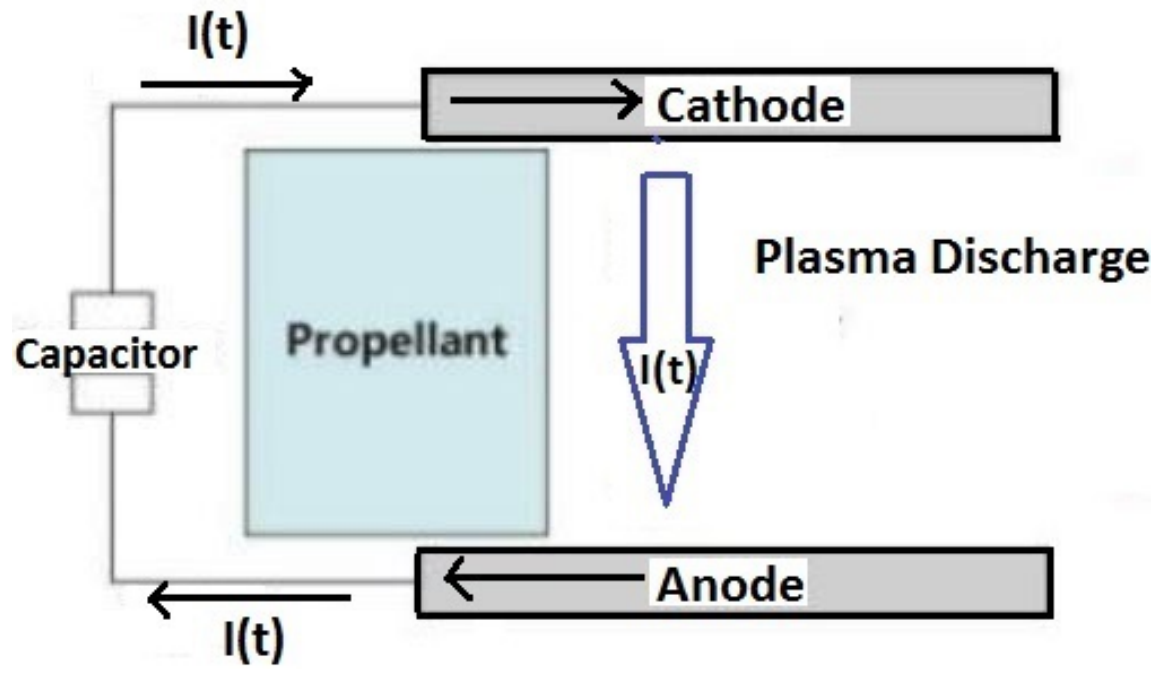

Figure 1: Schematic of an ablative PPT

A significant amount of attention has been given in the past to model ablative PPTs. Two different approaches have been used to formulate theoretical models for PPTs, 1) Simple 1-D models [2][14], 2) advanced MHD modelings [15][16]. One-dimensional models such as the slug/snowplow models combine the discharge electrical circuit equations with dynamical equations to provide the macro details behind the acceleration processes in PPTs. However, conventional slug/snowplow models are developed based on two assumptions which tend to deviate from real thrusters, 1) They assume complete ionization of the propellant mass, 2) They assume the existence of a single plasma sheet. Moreover, conventional 1-D models are dependent on the experimental results for the values of the total mass shot per discharge, which are then used to solve the lumped circuit equations. Conventional models have been recently modified to calculate the plasma mass shot [17][18]. However, the models don't account for the development of multiple plasma sheets, the presence of which is verified experimentally. Very recently, a two stream numerical model for ablative PPT was proposed [19], however, the model assumed multiple plasma sheets in parallel to the external circuitry, this method deviates from the experimental observations which suggests that the initial accelerating plasma sheet short circuits from the external energy supply once a new plasma sheet is generated [2]. A modified slug model accounting for multiple plasma mass was developed [20], however, the model assumes a constant and uniform plasma mass and depends on the experimental iteration values of the total mass shot per discharge to calculate the mass of individual plasma streams. A multiple plasma stream model was also presented by Nawaz et al [21]. The details of the model description and methods however, were very brief, additionally, the model utilized a constant individual plasma mass, the model was also dependent on the experimental parameters as an input. Recently, much advanced models have been proposed [22][23][24][25][26], but the complexity and the difficulty in the implementation of such models has also increased substantially.

Magneto-hydro-dynamics modelings (MHD modelings) for plasma acceleration processes in PPTs have been greatly modified. MHD modelings have been developed to account for the late-time ablation processes [18][27]. Moreover, a multiple plasma sheet model has been proposed recently [28], however, the model is based on the assumption that the plasma flow in the discharge chamber is discontinuous, additionally, for convenience the model assumes that two plasma streams collide and fuse together to form a single plasma stream. The major problem with MHD modelings is the complexity and difficulty in implementation of MHD models. MHD modelings for ablation fed PPTs were also developed assuming a one-dimensional quasi-steady mode of acceleration and supersonic expansions in the limit of high Reynold number [29]. Numerical calculations were also performed for ablation controlled discharges and were applied to the plasma calculations for a Micro-PPT [30][31]. A physicomathematical model for ablative PPTs was proposed to overcome the complexity of MHD codes [32]. Numerical models accounting for crowbar resistances in the equivalent L-C-R circuit equations have been proposed [33][34], however, the existence of multiple plasma streams have not been accounted for in these models.

The goal of the present paper is to develop a simple and an easy to implement multiple plasma stream numerical model for ablative pulsed plasma thrusters. The model should account for most of the critical experimental observa-tions made during the operation of PPTs. The primary aim of the model is to make it dependent on minimal number of experimental parameters as an input. The model would be implemented in MATLAB/SIMULINK environment.

## 2. Theoretical Model Description

The entire theoretical model is discussed and developed in the following subsections-1) Multiple plasma sheet mass model, 2) Electric circuit equations, 3) Plasma mass and dynamical model, 4) Neutral gas dynamical model.

### 2.1. *Multiple plasma sheet mass model*

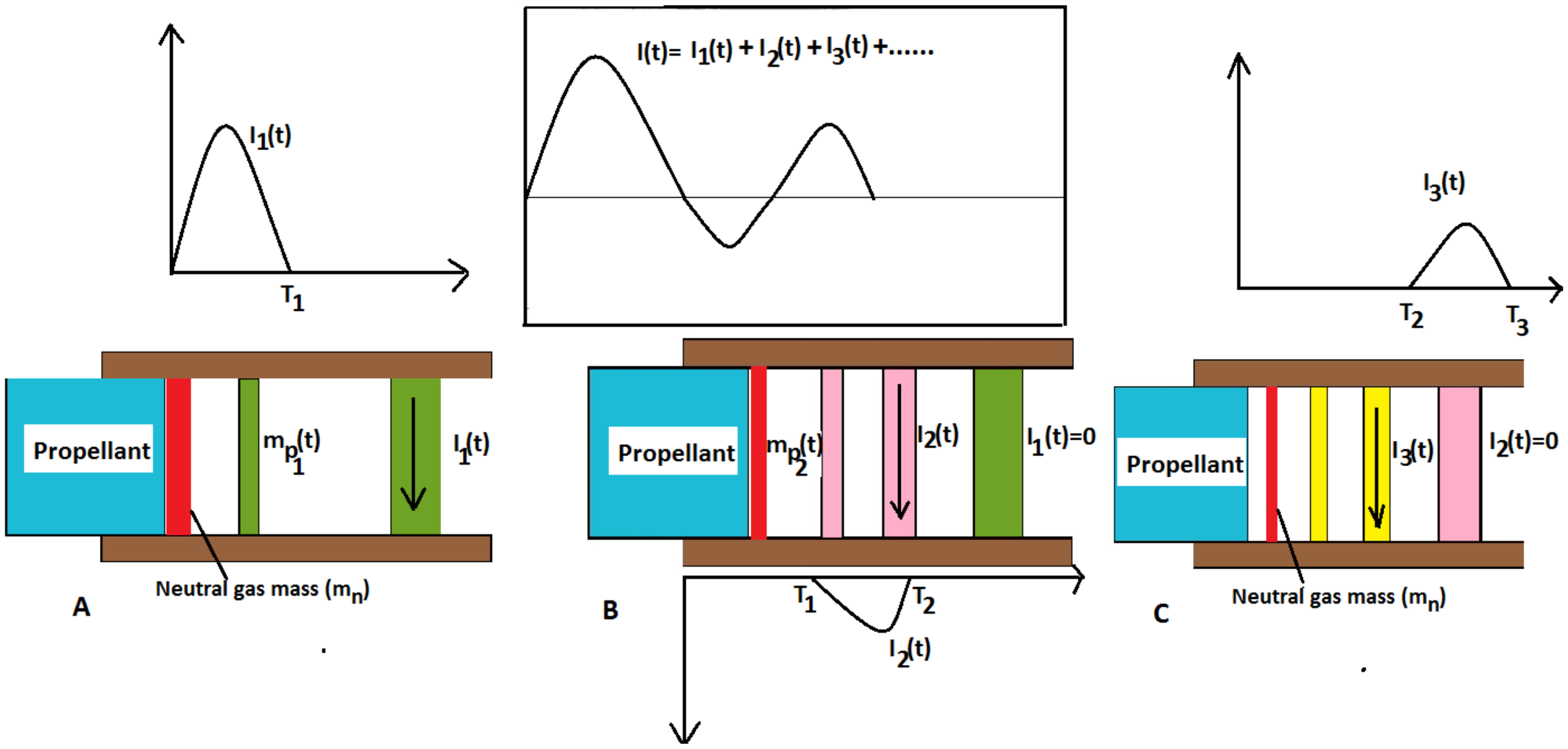

Figure 2: Multiple plasma sheet assumption

Experimental results have shown the presence of multiple plasmoids during the operation of PPTs [6][11]. Therefore, to develop a more general model we assume a multiple plasma sheet model. Figure 2 depicts the assumed plasma flow behavior for a multi-sheet model.

As soon as a discharge is initiated between the electrodes, the high temperature of the discharge arc heats the solid propellant surface, raising the surface temperature above the decomposition temperature of the propellant. This generates the neutral gas near the ablating surface. As the temperature raises further, a fraction of neutral gas is converted into the first plasma stream as shown in figure 2 (A). The neutral mass moves with a relatively much slower velocity when compared to the high plasma mass velocity. Although, it is not possible to accurately determine the time when the second plasma stream is formed, it is observed to form at a time slightly prior to the current reversal time. Therefore, for the present model we assume that the new (secondary) plasma sheet is formed when the current crosses a zero value. It is assumed that as soon as a discharge is initiated a single plasma stream carrying current ($I_1(t)$) exists during the first half cycle. The mass of the first plasma stream is denoted by ($m_{p1}(t)$). When the first half of the current cycle reaches zero at time ($T_1$), a fraction of the slow moving neutral gas is again ionized and a second plasma stream carrying a current and mass of ($I_2(t)$) and ($m_{p2}(t)$) respectively, is created as shown in figure 2 (B). It must be noted the second plasma stream is not formed at the surface of the propellant but rather at the position where the neutral gas is present during the current reversal time [13]. As soon as this second plasma sheet is formed, the first plasma sheet short circuits with the external voltage supply i.e. ($I_1(t) = 0$ for the second half cycle). The same process continues for each zero crossing of the discharge current as shown in figure 2. Therefore, the total discharge current is assumed to be discontinuous and calculated by superimposing current streams from different plasma sheets.

Therefore, we can represent the total discharge current ($I(t)$) as the sum of individual currents ($I_i(t)$) generated per cycle.

$$I(t) = \sum_{i=1}^{n} I_i(t) \tag{1}$$

where $I_i(t)$ represents the discharge current flowing through the i'th plasma sheet. The total plasma mass shot per discharge, could be similarly represented as the sum of each individual plasma streams mass,

$$m_p(t) = \sum_{i=1}^{n} m_{p_i}(t) \tag{2}$$

where, $m_{p_i}(t)$ is the mass of the i'th plasma stream. The i'th plasma stream is accelerated between the time interval ($t = T_{i-1}$ to $t = T_i$). Here $T_i$ represents the time when the i'th plasma stream is formed.

### 2.2. Electric circuit equation

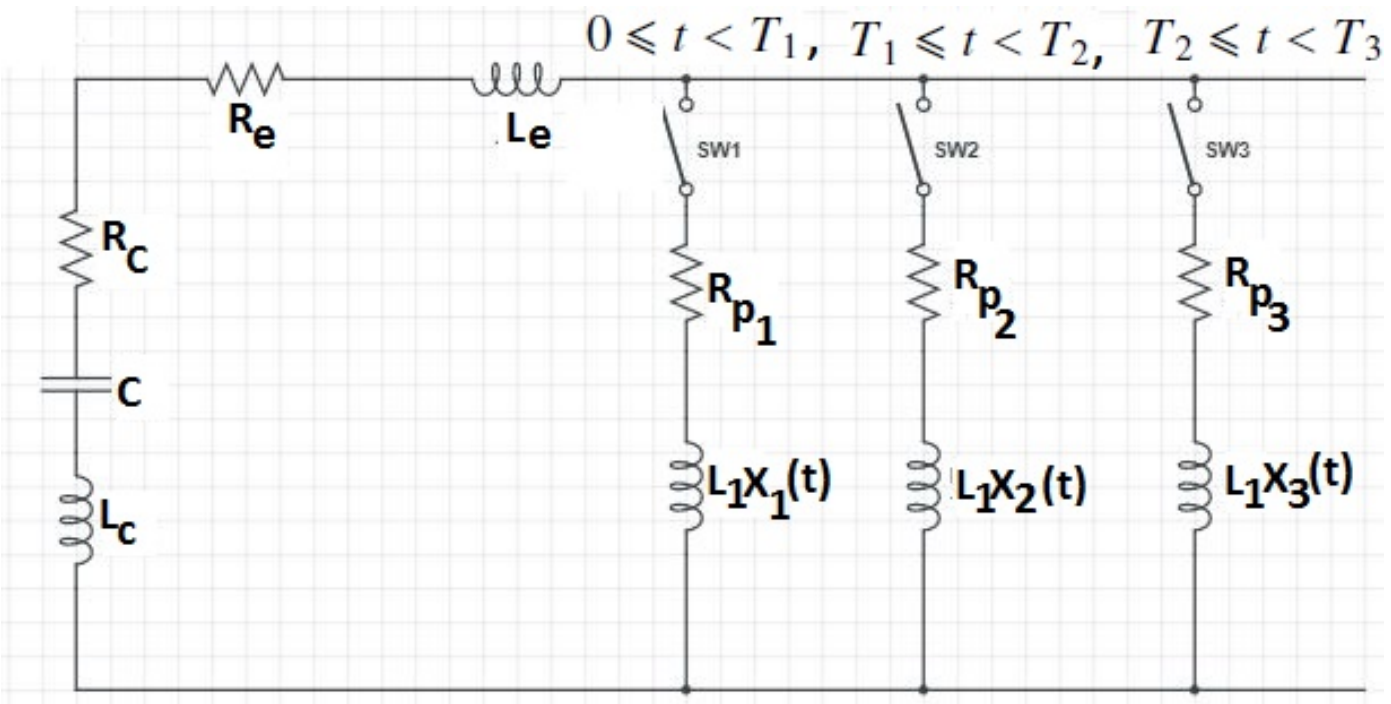

Figure 3: Equivalent Pspice electrical circuit for multiple plasma stream model

It is possible to describe the entire plasma acceleration process using an equivalent series L-C-R circuit. By combining the conventional slug model lumped circuit equations with the multiple plasma sheet assumption made in the paper, we assume that each i'th plasma stream carries a specific current $I_i(t)$ for the interval $T_{i-1}$ to $T_i$. The resistance and inductance of this i'th plasma stream is represented by $R_{p_i}$ and $L_1X_i(t)$. Where $L_1$ is the inductance per unit length for the given geometry of the thruster.

The equivalent L-C-R circuit for multiple plasma assumption is depicted in figure 3.

Therefore, applying the Kirchoff's law for the i'th plasma stream (that is when the i'th switch is closed) we get,

$$V_{i-1} - \frac{1}{C}\int_{T_{i-1}}^{t} (I_i(t))^2 dt = I_i(t)[R_c + R_e + R_{p_i}] + [L_c + L_e + L_1X_i(t)]\frac{dI_i(t)}{dt} + L_1 I_i(t)\frac{dX_i}{dt} \tag{3}$$

Where $V_{i-1}$ represents the voltage across the capacitor at time $t = T_{i-1}$. Here, $V_o$ represents the initial discharge voltage, $R_c$, $L_c$, $R_e$, $L_e$ are the resistances and inductances of the capacitor and transmission lines, respectively. $X_i(t)$ is the position of the i'th plasma stream in the discharge chamber.

The plasma inductance is a function of the position of the plasma sheet. Depending on the rectangular geometric configuration of the thruster several numerical model for the inductance per unit length of the rectangular geometry exist [2][1][14]. Most of the conventional numerical models assume the rectangular geometry of the thruster to be quasi-infinite in width ($w >> h$). Theses models tend to ignore the "Fringe" effects due to quasi-infinite assumption, therefore, the inductance-per-unit length is given by,

$$L_1 = \mu_o \frac{h}{w} \tag{4}$$

However, most practical PPTs don't follow the quasi-infinite electrode geometry configuration, therefore, a more accurate model as proposed by Burton et al. [1] is used in the present paper. The model incorporates the fringing effects in real PPTs. The equation using the modified model is thus given by [1],

$$L_1 = \frac{\mu_o}{\pi}[1.5 + ln(\frac{h}{w})] \tag{5}$$

The time varying nature of the plasma resistance is hard to predict numerically. No single accurate numerical model exists to predict the plasma resistance. Therefore, for the present model we use a constant plasma resistance value as proposed by Turchi et al. in their MACH2 codes for ablative PPTs [29]. The model is based on a quasi-steady assumption and in the limits of high magnetic Reynolds number. For an oscillatory current waveform behavior, the model would be in quasi-steady mode if the successive pulse timing is much greater than the total transit timing. The transit timing could be obtained by assuming a reasonable exit speed and dividing it by the electrode length. Therefore, using the model the total plasma resistance at the surface of the ablating surface could be given by [29],

$$R_p = \mu_o V_{crit} \frac{h}{w} \tag{6}$$

This equation reduces from the Faraday's law for steady state, plasma accelerator flow. Here, $V_{crit}$ is the Alfven critical speed, and depends on the degree of ionization and the choice of the propellant. For the model it is assumed that the plasma resistance remains constant throughout the acceleration process.

### *2.3. Plasma Mass and Dynamical Model*

The total mass shot per discharge is decomposed into fast moving plasma and slow moving neutral gas. Although it is not possible to accurately describe analytically the mass of the neutral gas generated per discharge, models for plasma mass generated per discharge do exist [15][18].

The plasma mass model for individual plasma sheet is derived from existing numerical model proposed by Henrikson et al [18]. The model is derived assuming a quasi-steady flow. For a magnetohydrodynamics flow under a high magnetic Reynolds number and the magnetic pressure much greater than the plasma pressure, there exits a magneto-sonic point. The conditions at this magnetic-sonic point is then used to derive the mass of the plasma sheet [18]. This model is used to evaluate the mass of individual plasma sheet, given by the relation below,

$$m_{p_i}(t) = \frac{\mu_o h}{w V_{crit} 4.404} \int_{T_{i-1}}^{t} I_i^2(t) dt \tag{7}$$

The plasma mass, depending on the mode of operation is only a fraction of the total mass shot. The remaining mass is decomposed into neutral mass due to the surface temperature of the propellant being greater then the decomposition temperature however, being less than the temperature to ionize the gas.

The force on the individual plasma sheet is assumed to be completely electromagnetic. The total force on the i'th plasma stream could be given by the Lorentz Force [2],

$$F_{L_i}(t) = \iiint \vec{j_i} X \vec{B_i}(x, y, z) dx dy dz \tag{8}$$

where, $\vec{j_i}$ represents the current density of the i'th plasma sheet and $\vec{B_i}(x, y, z)$ represents the self magnetic field. Equation 8 could be greatly simplified in terms of the discharge current and inductance-per-unit length terms as given below,

$$F_{L_i}(t) = \frac{1}{2} L_1 I_i^2(t) \tag{9}$$

The total force on the i'th plasma stream could be represented in terms of the mass and position of the i'th sheet in the discharge chamber, therefore,

$$\frac{d}{dt}[m_{p_i}(t) \frac{dX_i(t)}{dt}] = \frac{1}{2} L_1 I_i^2(t) \tag{10}$$

Combining equation 7 and 10 we can conclude the plasma dynamics behavior for the i'th plasma stream by the following differential equation relation,

$$\frac{\mu_o h}{w V_{crit} 4.404} I_i^2(t) \frac{dX_i(t)}{dt} + \frac{d^2 X_i(t)}{dt^2} [\frac{\mu_o h}{w V_{crit} 4.404} \int_{T_{i-1}}^{t} I_i^2(t) dt] = \frac{1}{2} L_1 I_i^2(t) \tag{11}$$

It must be noted that equation 11 is a function of the i'th plasma current and the position of the i'th plasma stream.

### *2.4. Neutral gas dynamical model*

Unlike the plasma model description, neutral gas mass behavior is hard to describe analytically and none of the existing models are accurate. Since neutral gas particles are slow moving in nature with the mean exit velocity times less than the plasma velocity, neutral gas mass has no significant impact on the net thrust delivered. However, due to the heavier nature of neutral sheets the propellant utilization of the thruster degrades [8]. For the present section we assume that the neutral mass similar to the plasma sheet assumption is a uniform rectangular sheet. As pointed out in section 2.1, the neutral gas is initially formed at the propellant surface, a fraction of which is ionized and forms the first plasma stream and accelerated till the first half cycle (till t=$T_1$), following this a second plasma stream is formed at the position of the neutral gas at the first current reversal time ($T_1$). Therefore, to determine the initial position of the i'th plasma stream ($X_i(T_i - 1)$), the position of the neutral sheet at time (t=$T_{i-1}$) must be known.

Although no accurate model exists for the decomposed neutral mass, Mikellides et al. [15], proposed a model to calculate the velocity of the neutral mass sheet. The model assumes a constant velocity for the neutral gas which is consistent with the experimental observations [13]. For a choked neutral gas flow, the velocity could be given by the relation,

$$V_{dec} = \frac{2}{9} \frac{B_o^2}{\mu_o V_{crit}} [\frac{R T_{dec}}{p_{eq}(T_{dec})} - \frac{1}{\rho_s}] \tag{12}$$

where, $R$, $T_{dec}$, $\rho_s$, $p_{eq}$ are the specific gas constant (For Teflon monomer $R = 83.14 K/Kg/K$), decomposition temperature, density of solid and equilibrium vapor pressure[3]. It must be noted that equation 12 gives the upper limit of the decomposed (neutral) mass.

The position of the neutral gas sheet with respect to the propellant surface could be thus given by,

$$X_{dec}(t) = V_{dec} t \tag{13}$$

Therefore, equation 12 and 13 are combined to predict the position of the neutral layer. The initial position of the i'th plasma stream could be given by,

$$X_i(T_i - 1) = V_{dec} T_{i-1} \tag{14}$$

Since no accurate mass model for neutral gas exists at this point, the neutral gas mass is evaluated by comparing the calculated total plasma mass shot per discharge with the total mass shot (determined experimentally). This value of the neutral gas represents all the neutral gas which was accelerated per discharge (including the mass generated by the process of late time ablation).

## 3. Theoretical Model Implementation

The differential equations discussed in the last section is now summarized and implemented in a MATLAB/SIMULINK environment. We first start with combining the circuit equations with the dynamical equation.

The discharge current and position of the i'th plasma stream could be calculated by solving the following pair of differential equations,

---

[3] For discussion about the calculation for equilibrium vapor pressure for Teflon, refer to Appendix C of [15]

$$V_{i-1} - \frac{1}{C}\int_{T_{i-1}}^{t}(I_i(t))^2 dt = I_i(t)[R_c + R_e + R_{p_i}] + [L_c + L_e + L_1 X_i(t)]\frac{dI_i(t)}{dt} + L_1 I_i(t)\frac{dX_i}{dt};$$
$$\frac{\mu_o h}{wV_{crit}4.404} I_i^2(t)\frac{dX_i(t)}{dt} + \frac{d^2X_i(t)}{dt^2}[\frac{\mu_o h}{wV_{crit}4.404}\int_{T_{i-1}}^{t} I_i^2(t)dt] = \frac{1}{2}L_1 I_i^2(t) \quad (15)$$

The initial position for the i'th plasma sheet and plasma mass is evaluated using the following relation,

$$X_i(T_i - 1) = V_{dec}T_{i-1} \quad \text{and} \quad m_{p_i}(t) = \frac{\mu_o h}{wV_{crit}4.404}\int_{T_{i-1}}^{t} I_i^2(t)dt \quad (16)$$

The pair of differential equation given by relation 15 is solved for individual plasma stream starting with the first plasma stream. The results are then superimposed to calculate the total discharge current and plasma mass shot. The position of i'th plasma stream is set using relation 16 as the initial condition for each plasma sheet given in equation 15. Equation 15 is first solved for the first plasma stream to obtain the first half current cycle and plasma mass shot. Zero crossing detectors are used in the SIMULINK model to obtain the value of $T_1$ (Note that $T_0 = 0$). The position of the neutral gas at this time is then calculated which becomes the initial position of the second plasma stream. The same process is extended till the capacitor is completely drained.

The total discharge current and plasma mass shot is therefore, given by,

$$I(t) = \sum_{i=1}^{n} I_i(t) \quad \text{and} \quad m_p(t) = \sum_{i=1}^{n} m_{p_i}(t) \quad (17)$$

The total plasma mass shot is then compared with the experimental results for the total mass shot to obtain the values for the neutral gas mass produced per discharge.

## 4. Results and Discussions

In the present section we validate and discuss the entire model formulated in the previous sections. First we validate the model for a well studied ablative thruster developed at the University of Tokyo [6][13], for simplicity we name this thruster as APPT-1. Next we validate and discuss the performance for the Lincoln Experimental Satellite thruster (LES-6) [9], we name this thruster as APPT-2 in the present paper. In the last subsection we make an attempt to extend the model and test it for a liquid PPT developed at the University of Tokyo, for convenience we name this thruster as LPPT [13].

### *4.1. Discussion for APPT-1*

The model is first vaidated and discussed for an ablative PPT developed at the University of Tokyo [6][13].Table 1 depicts the operational parameters for APPT-1 [13].

Simulations were performed for first and second degree of ionization of the propellant, however, the simulation results for second degree of ionization were in good agreement with the experimental results. Therefore, simulation results only for this case is presented in the paper.

| **Operation parameter** | **Value** |
|---|---|
| Electrode height | $20mm$ |
| Electrode width | $10mm$ |
| Electrode length | $25mm$ |
| Capacitor (C) | $3\mu F$ |
| Discharge voltage ($V_o$) | $2200V$ |
| Total mass shot per discharge ($m_T$) | $7.1\mu g$ |

Table 1: Operation parameter for APPT-1

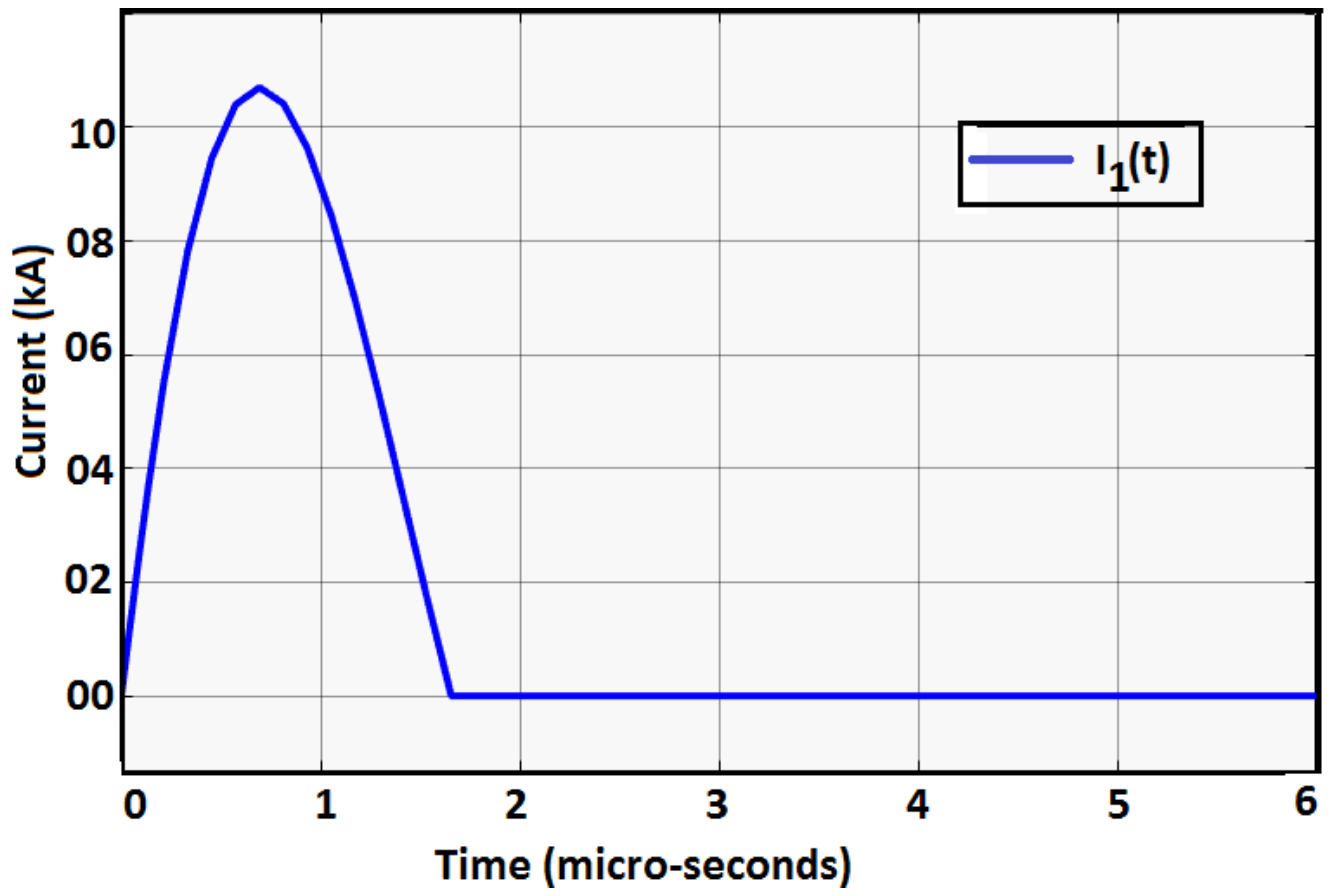

Figure 4: Simulation result of discharge current for first plasma stream for APPT-1

The simulation is started by calculating the dynamics of 1st plasma stream. In the circuit equations, the initial discharge voltage derived from the physical setup is used. Additionally, the transmission line resistance and inductance is assumed and set to zero. Initial time is set to $T_0 = 0$. The initial position position of the first plasma stream is taken to be the reference at $X = 0$. The inductance per unit length and plasma resistance calculated using relations 5 and 6 are $0.84\mu H/m$ and $48m\Omega$, respectively. Implementing the pair of differential equation 15 for the first plasma stream reveals the discharge current and position of the first plasma stream. The plot gives the value for the first zero ($T_1$), which is then further used to calculate the current and dynamics of subsequent plasma stream. Figure 4 represents the simulation result for discharge current of the first plasma stream for APPT-1.

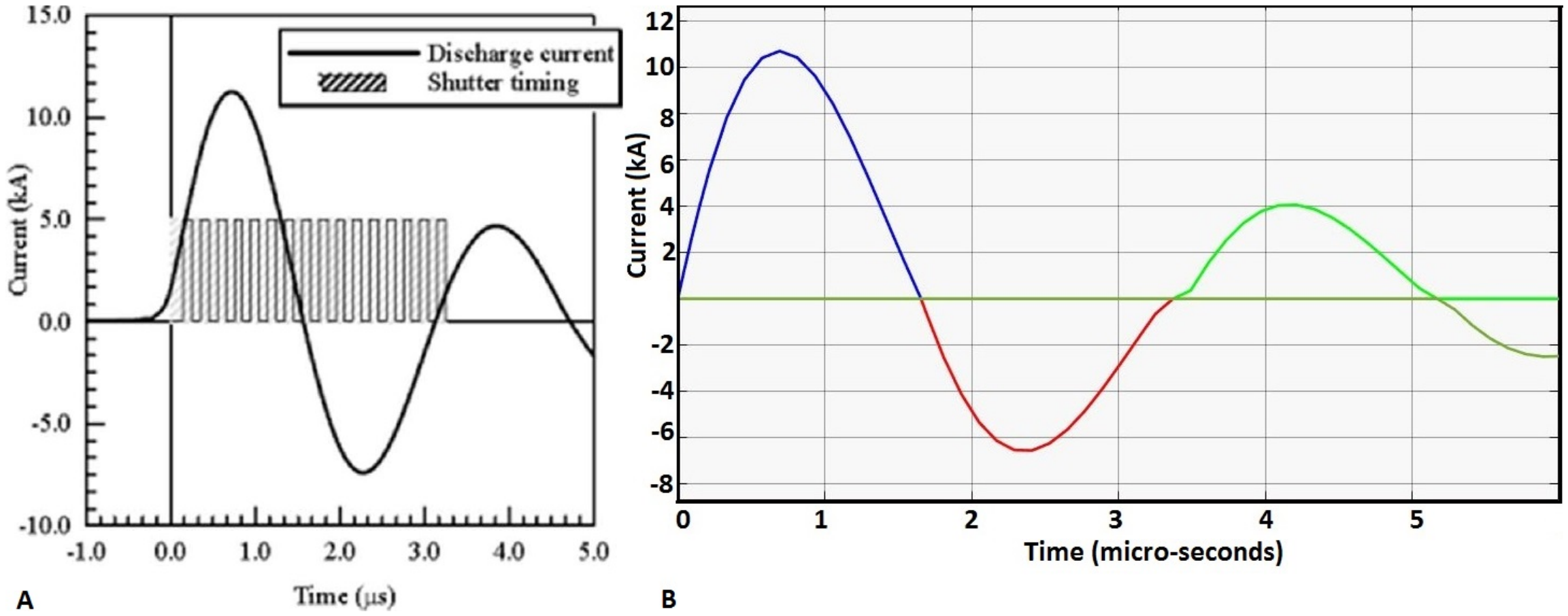

Figure 5: Discharge current for APPT-1 A) Experimental result [13], B) Simulation result

Simulations were performed for the first four plasma streams, subsequent plasma streams won't have any noticeable contribution to the total thrust performance due to the relatively low mass. By superimposing the simulation results for the initial four plasma stream , the discharge current predicted by the model is obtained and is compared with the experimental values for APPT-1 as shown in figure 5.

The experimental plot and the simulation results are in good agreement. The experimental results for APPT-1 suggests the first cycle lasts till $t = 1.57\mu s$, whereas, the simulations suggest ($T_1$) at $1.63\mu s$. The peak currents are in good agreement. For the second cycle, the initial position of the neutral mass stream is used as the initial condition for the plasma stream. For each zero crossing time, the simulated current waveform is slightly discontinuous, this is due to the setting of the initial secondary plasma velocity back to zero abruptly at the current reversal. Therefore, there is an abrupt change to the total circuit inductance.

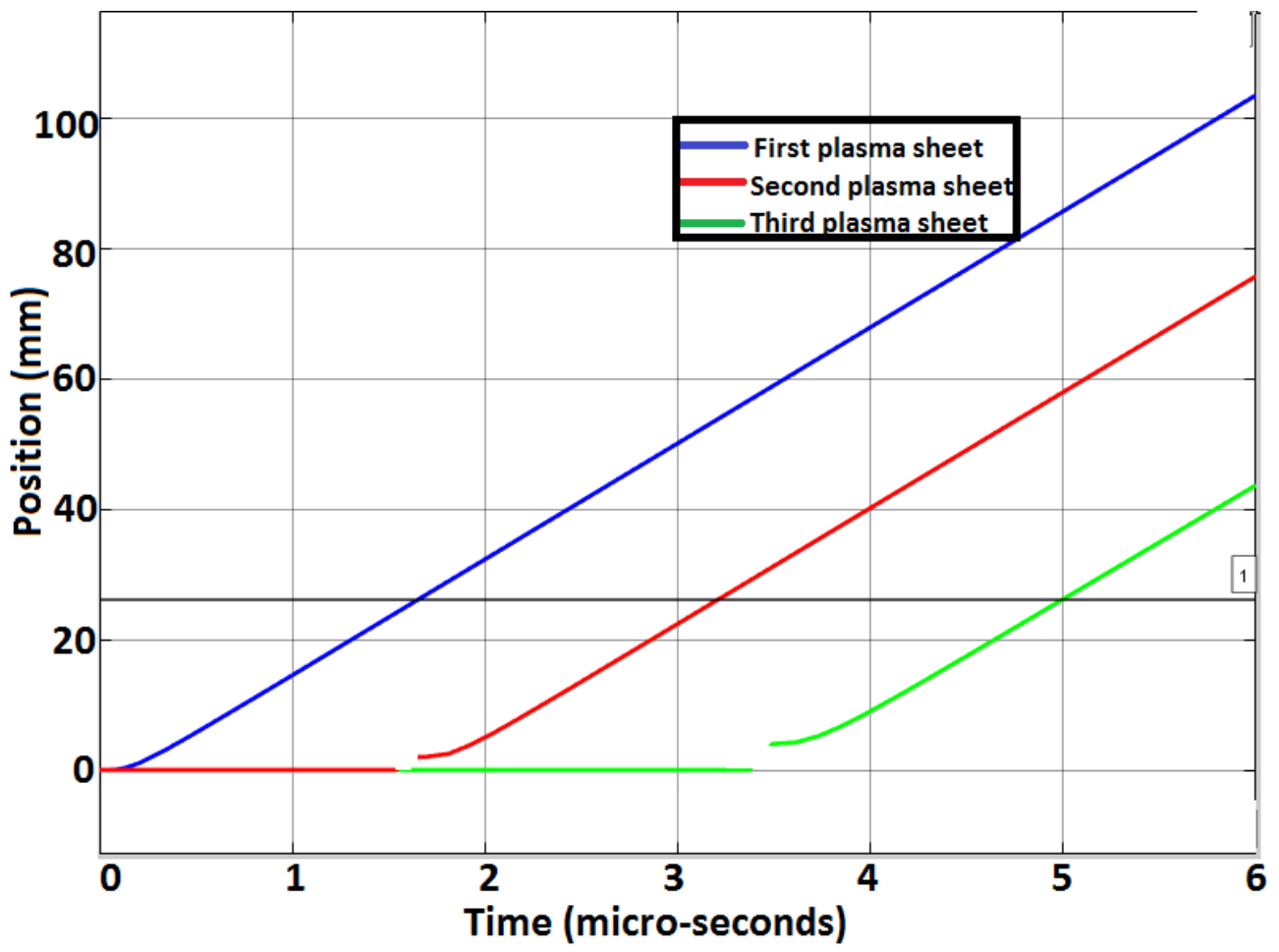

Figure 6: Simulation for position of plasma stream with respect to time

The plots of the position of the first three plasma streams with respect to time is depicted in figure 6. The mean exit velocity of the first plasma stream suggested by the simulation is close to $17.8Km/s$. This is in close agreement with the experimental results of $16.1Km/s$. The mean exit velocity of the second and third plasma stream was predicted to be $17.2Km/s$ and $17.1Km/s$, respectively. The plasma mass and the impulse bit delivered by respective plasma sheets is depicted in figure 7. Practically speaking, after the initial three plasma streams, the electrical energy from the capacitor won't be sufficient enough to generate any subsequent plasma streams, however, for evaluating the continuous discharge current, the model assumes the existence of plasma in the discharge chamber till the capacitor is completely discharged.

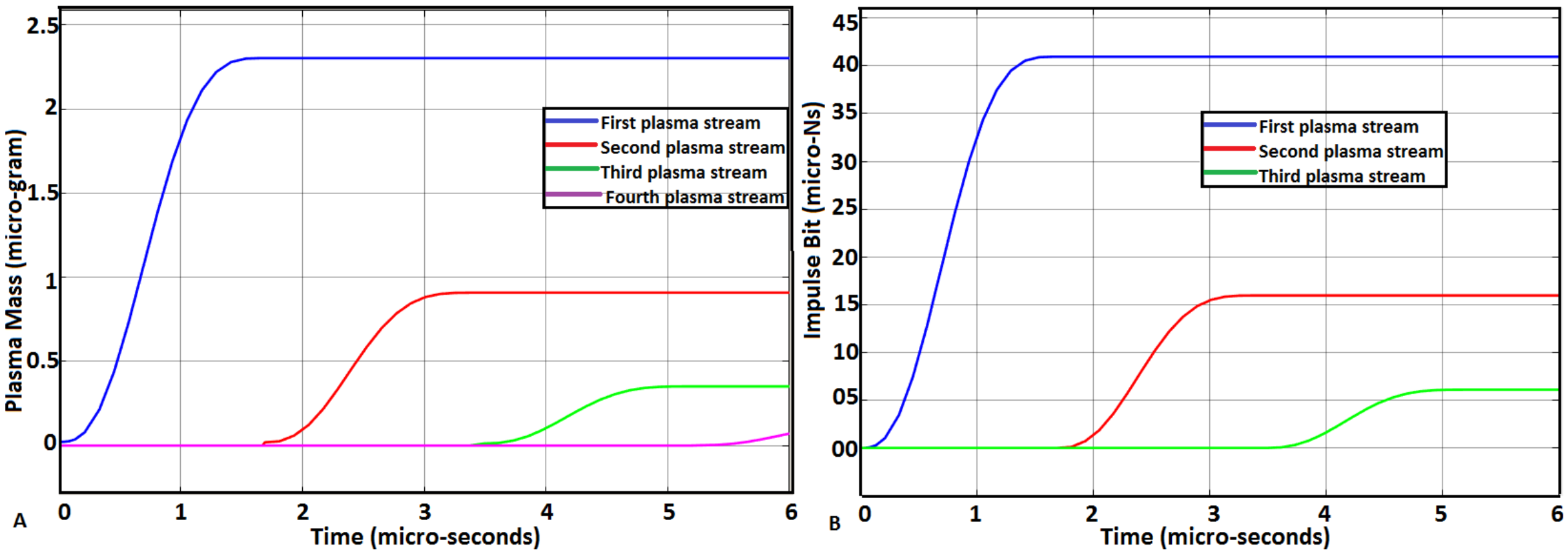

Figure 7: Simulation result for APPT-1 A) Plasma mass shot, B) Impulse bit delivered

Simulations suggest that only $3.8\mu g$ of the total supplied mass of $7.1\mu g$ forms the plasma stream. The remaining mass is not ionized and forms the slow moving neutral layer of mass. This leads to the poor propellant utilization. The mass of the first plasma layer is the greatest, that is 62*percent* of the total plasma mass shot, and hence has the most significant contribution to the net thrust performance. The total impulse bit delivered per discharge by the plasma is calculated to be $63.7\mu - Ns$. The model slightly over-predicts the total impulse bit when compared to the experimental results of $55\mu{-}Ns$. Table 2 compares the predictions made by the numerical model with the experimental results. The difference in the total mass shot between the experimental result and simulation result clearly shows that a great fraction of mass bit is not ionized.

| **Parameter** | **Experimental** | **Simulation** | | |
|---|---|---|---|---|
| | | 1st plasma | 2nd plasma | 3rd plasma |
| Current cycle period ($\mu s$) | 1.57 | 1.63 | 1.61 | 1.63 |
| Mass Bit ($\mu g$) | 7.1 | 2.32 | 0.91 | 0.38 |
| Impulse Bit ($\mu N - s$) | 55 | 40.9 | 15.8 | 5.96 |
| Mean Plasma Exit Velocity (m/s) | 15700-16100 | 17800 | 17200 | 17100 |
| Specific Impulse ($s$) | 790 | 1816 | 1755.10 | 1744.89 |

Table 2: Comparison of performance parameters for APPT-1

### *4.2. Discussion for APPT-2*

LES-6 satellite thruster [7] has been studied very extensively in the past both experimentally and numerically. In the present section we validate the model for this thruster. Table 3 presents the operational parameters for the LES-6 thruster [9].

| **Operation parameter** | **Value** |
|---|---|
| Electrode height | $30mm$ |
| Electrode width | $10mm$ |
| Electrode lenght | $6mm$ |
| Capacitor (C) | $2\mu F$ |
| Discharge voltage ($V_o$) | $1360V$ |
| Total mass shot per discharge ($m_T$) | $10\mu g$ |

Table 3: Operation parameter for APPT-2

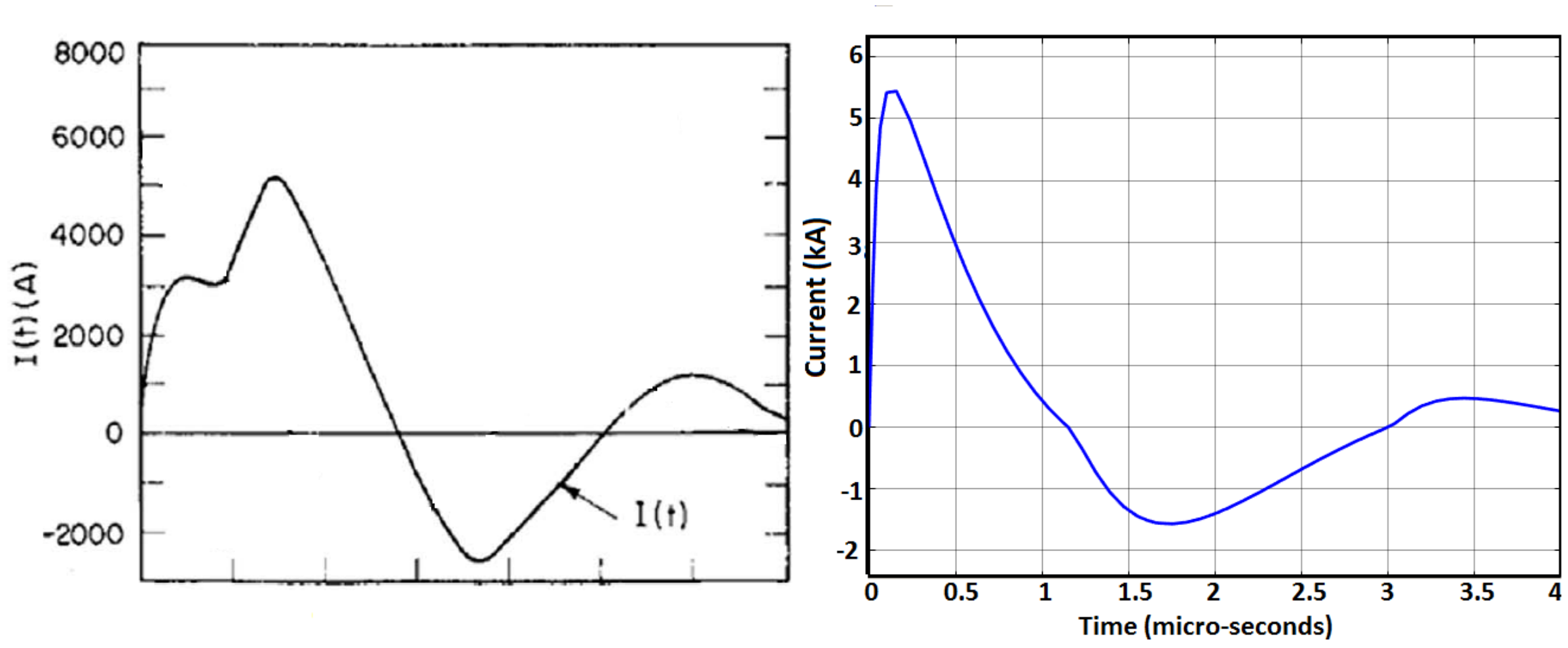

Figure 8: Discharge current for APPT-2 A) Experimental result [9], B) Simulation result

The simulated discharge current and it's comparison with the experimental result is depicted in figure 8. The plasma resistance was calculated to be $53m\Omega$. It could be observed that the simulation result for the discharge current is non differentiable for the time when zero crossing occurs. This could be explained due to the discontinuous current sheet assumption made by the model. As soon as the zero value is crossed by the discharge current, the initial position and velocity of the plasma stream is changed to initial value abruptly, therefore, for each time the discharge current crosses zero, the rate of change of the discharge current changes abruptly for that value. This change was not observed to be very abrupt for APPT1 as due to the operational configuration the first plasma stream left the discharge chamber as soon as the current reversal occurred. Therefore, the total circuit inductance didn't reach an unreasonably high value before being initialized back to initial value.

First current reversal as suggested by the experimental results happens at $t = 1.1\mu s$, the model suggests the first reversal to take place at $T_1 = 1.24\mu s$. Simulations were performed for the initial three plasma streams.

The neutral sheet mass velocity was calculated to be close to $720m/s$. Therefore, the initial position of the i'th plasma stream is determined using the current reversal time and constant velocity of the neutral mass stream. Figure 9 depicts the velocity distribution for the first three plasma streams. The model predicts the mean exit velocity of first, second and third plasma stream as $29Km/s$, $27.9Km/s$ and $28.2Km/s$, respectively. The experimental results suggest a mean exit velocity of the ions to be close to $40Km/s$. From simulations it is evident that the neutral mass velocity is significantly lower when compared to the velocity of the plasma sheets.

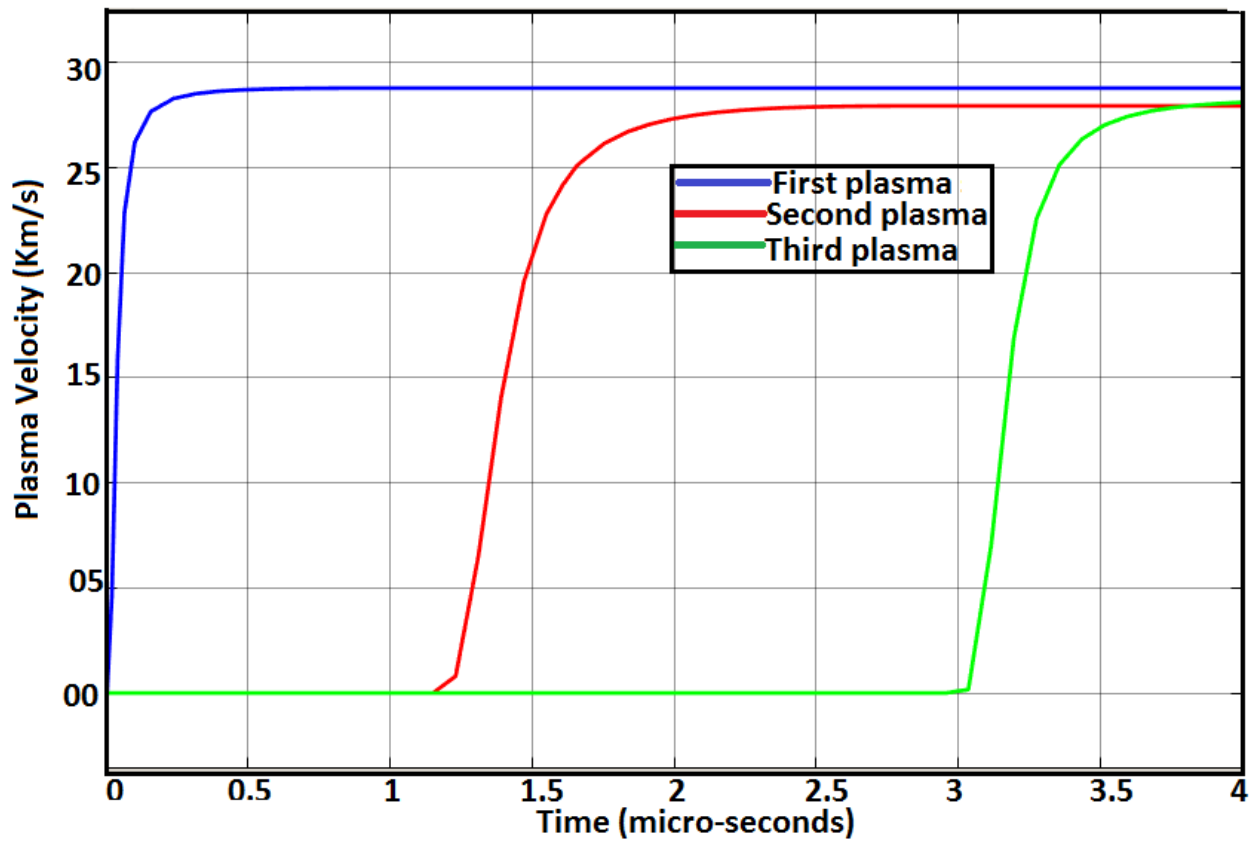

Figure 9: Simulation for plasma velocity for APPT-2

The discharge current evaluated is used to simulate the individual plasma mass and the impulse bit delivered by individual plasma stream. The variation in the mass of individual plasma stream is depicted in figure 10 (A), the first plasma stream carries the majority of the mass (and hence the momentum), however, the total plasma shot calculated by the simulation results is close to $0.9\mu g$. The total ionized mass predicted by the experimental results is close to $1\mu g$. Therefore, the model accurately predicts the plasma mass shot per discharge. Figure 10 (B) depicts the simulations for the total impulse bit delivered. Simulation suggests the total impulse bit delivered by the plasma streams is close to $27\mu N - s$, experimental results suggest a value of $31.2\mu N - s$, therefore, the remaining impulse is generated by the neutral mass stream.

From the analysis it could be concluded that despite the heavy mass of the neutral sheet (approximately 90 percent of the total mass shot), the total impact on the net thrust performance is significantly low. This is the primary reason behind the low efficiency observed for APPTs.

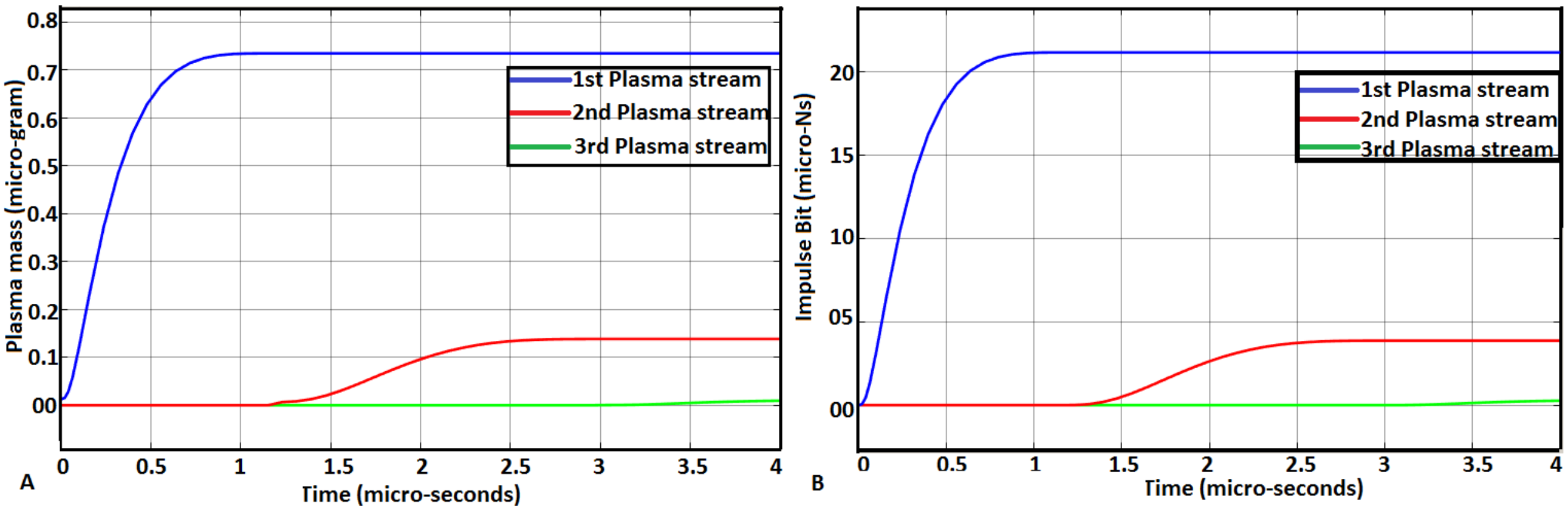

Figure 10: Simulation result for APPT-2 A) plasma mass shot B) Impulse bit

Table 4 summarizes the comparison between the simulation results and the experimental results.

| **Parameter** | **Experimental** | **Simulation** | | |
|---|---|---|---|---|
| | | 1st plasma | 2nd plasma | 3rd plasma |
| Current cycle period ($\mu s$) | 1.1 | 1.24 | 1.72 | 1.3 |
| Plasma mass shot ($\mu g$) | 1 | 0.73 | 0.14 | 0.02 |
| Impulse Bit ($\mu N - s$) | 31.2 | 22.1 | 4.3 | 0.3 |
| Mean Plasma Exit Velocity (m/s) | 40000 | 29000 | 27900 | 28200 |
| Specific Impulse ($s$) | 304.3 | 2959 | 2846 | 2877 |

Table 4: Comparison of performance parameter for APPT-2

### *4.3. Discussion for LPPT*

We now make an attempt to extend the numerical model developed in present paper for liquid propellant PPTs. For liquid propellant thrusters fast acting injectors are used for feeding the propellant into the discharge chamber, this makes the total mass shot per discharge controllable. Therefore, a more rigorous mass ablation model needs to be developed, for liquid fed PPTs, however, the multiple plasma sheet assumption and the circuit equations would still hold valid and meaningful. The goal of this section is to validate the plasma multiple sheet model and the equivalent electrical circuit model. We test a $7.5J$ class of thruster developed at University of Tokyo, the operational parameters of LPPT is similar to APPT-1 discussed before. Table 5 presents the operational parameters for LPPT [13].

| **Operation parameter** | **Value** |
|---|---|
| Electrode height | $20mm$ |
| Electrode width | $10mm$ |
| Electrode lenght | $35mm$ |
| Capacitor (C) | $3\mu F$ |
| Discharge voltage ($V_o$) | $2200V$ |
| Total mass supplied per discharge ($m_T$) | $3\mu g$ |
| Total discharge resistance ($R_T$) | $64m\Omega$ |

Table 5: Operation parameter for LPPT

We model the discharge current for the initial three plasma streams and the simulation results and it's comparison with the experimental result is depicted in figure 11.

It could be observed the circuit equations for LPPT under-predicts the peak values for the discharge current when compared to the experimental data. The first peak current shown experimentally is $14kA$, whereas, simulations suggests a peak close to $9.5kA$. The time period of the discharge cycles calculated by the simulations are in good agreement with the experimental data. The first current reversal occurs at $1.60\mu s$ as pointed by the experimental results. Simulations suggest a value close to $1.76\mu s$. Simulation for the total plasma mass shot and the impulse bit delivered by LPPT is depicted in figure 12.

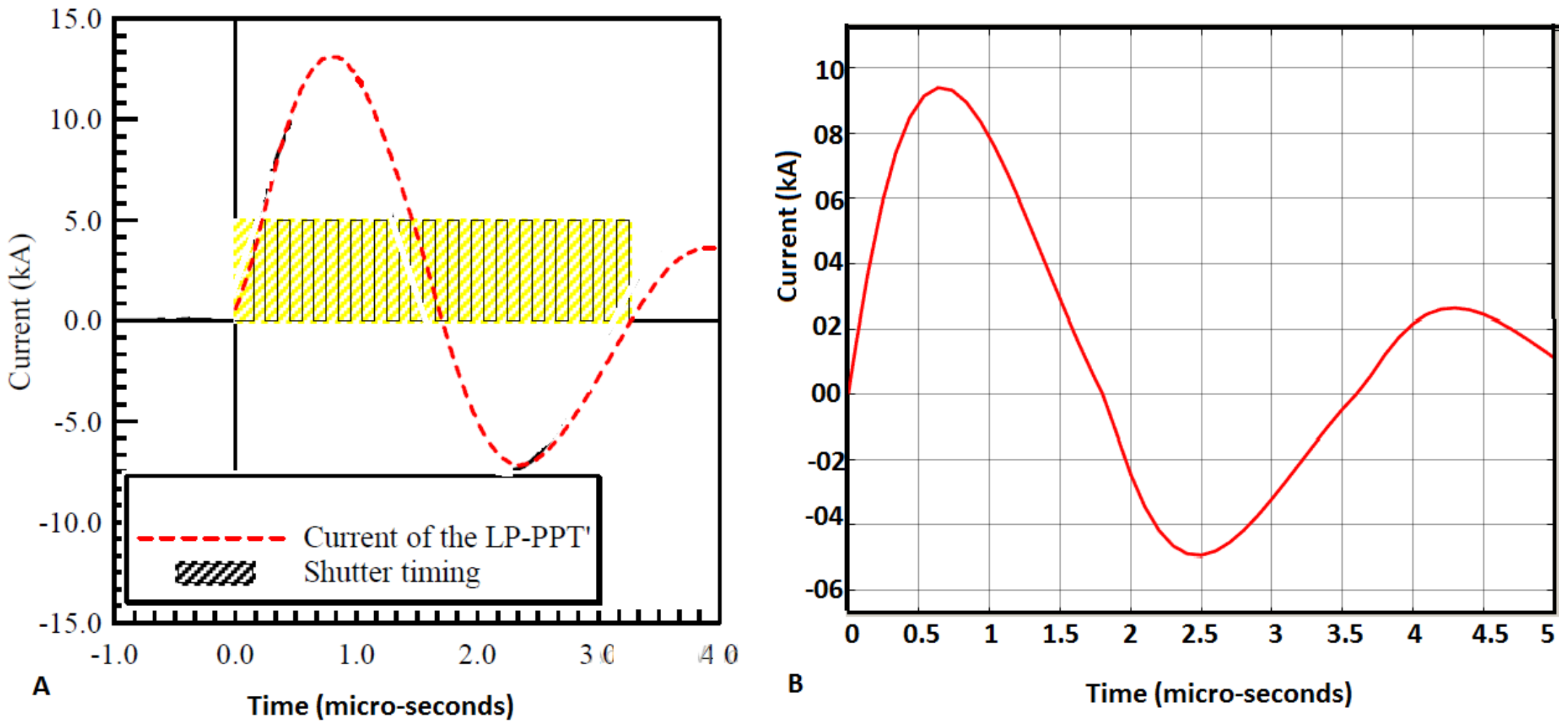

Figure 11: Discgharge current for LPPT A) Experimental result [13], B) Simulation result

The initial plasma stream carries a mass of $0.85\mu g$, followed by the second and third plasma stream of $0.23\mu g$ and $0.08\mu g$ respectively. The propellant utilization efficiency for liquid PPT is much better when compared to APPT. The plasma mass shot determined experimentally is $1.1\mu g$. Moreover, since injectors are used to control the mass shot per discharge, the propellant utilization could be further improved. The model over-predicts the total impulse bit delivered. Experiments suggest $37\mu Ns$ of total impulse is generated by LPPT. However simulation results suggest a value close to $48\mu Ns$.

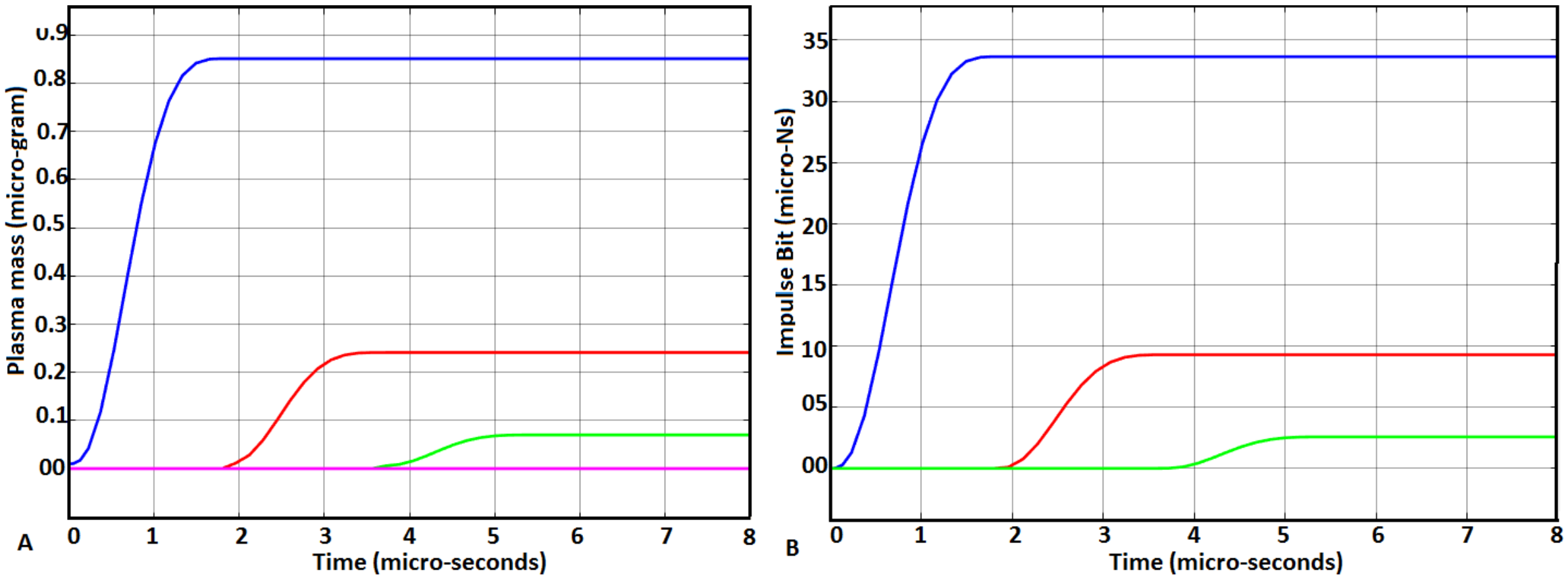

Figure 12: Simulation for LPPT A) Plasma mass shot, B) Impulse bit

Comparing the results for LPPT with the predictions made by the model for APPT, it could be concluded that the model is not very accurate for liquid propellant thrusters, however, it does provide a gross dynamical behavior of the individual plasma streams.

## 5. Conclusion

An one-dimensional multiple plasma stream numerical model is presented in this paper. The model is developed by combining the experimental observations of multiple secondary plasma streams with the existing numerical models. The model unlike the conventional slug/snowplow models doesn't rely on the experimental data to determine the dynamics of plasma sheets. In general the model is developed for ablative pulsed plasma thrusters. However, the plasma sheet model and the electrical circuit model could be used for liquid thrusters as well. From the simulation results it could be concluded that-

1) The model shows good accuracy for ablative pulsed plasma thrusters in predicting the dynamics of plasma sheets. Critical plasma parameters such as plasma position, velocity, plasma mass and the net impulse bit delivered could be accurately predicted using the model.

2) Simulations suggest only a small fraction of the total mass shot actually forms the plasma and accelerated to high exit velocity. The model could be further used to evaluate the variation in circuit and geometric parameters and it's impact on the plasma mass shot.

3) Further discussion is needed for the ablation and plasma creation processes for liquid pulsed plasma thrusters, however, the model is still valid to predict the motion and discharge current of the plasma stream. The model is useful and accurate to predict the plasma mass shot for liquid thrusters. Unlike ablative PPTs where the total mass shot is uncontrollable, liquid PPTs have injector based feeding mechanism. Therefore, the model could be used for careful design and study of high propellant efficiency liquid thrusters.